\documentclass[reviewcopy]{elsarticle}

\usepackage[reviewcopy]{adndt}
\usepackage{longtable}

\usepackage{mathptmx}

\usepackage{amsmath}

\usepackage{amssymb}

\biboptions{square,sort&compress}
\bibpunct[]{[}{]}{,}{n}{}{;}
\begin{document}

\begin{frontmatter}

\journal{Atomic Data and Nuclear Data Tables}

\title{Energy level properties of 4p$^6$4d$^3$, 
4p$^6$4d$^2$4f and 4p$^5$4d$^4$ configurations of W$^{35+}$ ion}

  \author{P. Bogdanovich\corref{cor1}}
 \ead{Pavelas.Bogdanovicius@tfai.vu.lt}

  \author{R. Kisielius}

  \cortext[cor1]{Corresponding author.}

   \address{Institute of Theoretical Physics and Astronomy, Vilnius University,
 A. Go{\v s}tauto 12, LT-01108 Vilnius, Lithuania}

\date{16/03/2012} 

\begin{abstract}  
The {\it ab initio} \/ quasirelativistic Hartree-Fock method developed 
specifically for the calculation of spectroscopic parameters of heavy atoms 
and highly charged ions was used to derive spectral data for the multicharged
tungsten ion W$^{35+}$. The configuration interaction method was applied to 
include the electron-correlation effects. The relativistic effects were taken 
into account in the Breit-Pauli approximation for quasirelativistic 
Hartree-Fock radial orbitals. The energy  level spectra, radiative lifetimes 
$\tau$, Lande $g$-factors are calculated for the $\mathrm{4p^64d^3}$, 
$\mathrm{4p^64d^24f}$ and  $\mathrm{4p^54d^4}$ configurations of the W$^{35+}$ 
ion. 

\end{abstract}

\end{frontmatter}

\newpage

\tableofcontents
\listofDtables
\listofDfigures

\section{Introduction}{\label{intro}}

In the present work we continue our theoretical studies \cite{01,02,03,04,05,06}
of multicharged tungsten ions.  The spectroscopic parameters of systems with the
open 4d-shell are investigated. Extensive interest in highly-charged tungsten 
ions \cite{07,08,09,10,11,12} is caused by its unique physical properties. 
Metallic tungsten is widely exploited in high-temperature devices, including 
fusion reactors \cite{13,14}. Although tungsten is difficult to melt and 
vaporize, its highly-charged ions emerge in fusion plasma and decrease its
temperature. Therefore one needs to control the concentration of these ions by
monitoring their spectra. Unfortunately, the experimental data for the
multicharged tungsten ions are very sparse. This fact is very evident from the 
recent compilation \cite{15} of the experimental and semi-empirical 
data for the tungsten ions. It is absolutely clear from that review that the 
tungsten ions from the rubidium-like to the palladium-like systems, which have 
an open 4d-shell, are investigated very seldom. For the most ions, only a few 
levels from the ground $\mathrm{4p^64d^N}$ configuration are determined. 
Moreover, very few energy levels of the first excited $\mathrm{4p^54d^{N+1}}$ 
and $\mathrm{4p^54d^{N-1}4f}$ configurations are determined for some tungsten 
ions. This situation has encouraged us to perform the theoretical investigation 
of the spectroscopic parameters for these ions. The calculated data can 
substantially facilitate experimental studies of the corresponding spectra; on 
the other hand, these data can be adopted directly for plasma spectra modelling.

The W$^{35+}$ ion energy spectrum of the $\mathrm{4p^64d^3}$, $\mathrm{4p^64d^24f}$ 
and $\mathrm{4p^54d^4}$ configurations consists of significantly larger number
of energy levels, and the range of possible transition wavelengths is wider
compared to the previously studied W$^{37+}$ \cite{05} and W$^{36+}$ \cite{06} 
ions. In order to keep the current paper concise, we present only the energy 
spectra and the characteristics of levels. The remaining spectroscopic 
parameters will be available from the database ADAMANT (Applicable DAta of 
Many-electron Atom eNergies and Transitions) which is currently under 
development. Likewise for the ions mentioned above ions, we employ a quasirelativistic 
(QR) approach \cite{16,17,18} with an extensive inclusion of correlation effects
to investigate the W$^{35+}$ ion spectroscopic properties. Since application of 
this approach has produced high-accuracy results for the W$^{37+}$ and W$^{36+}$
ions, there is no doubt about its applicability to the tungsten ion of slightly 
lower ionization degree.

In Section~\ref{calc} we provide a description of the adopted calculation 
method. Since this approximation ultimately matches the one applied for the 
investigation of the W$^{37+}$ and W$^{36+}$ spectroscopic data in 
\cite{05, 06}, here we provide only a brief summary of our method. Produced 
results are discussed in Section~\ref{result}.

\section{Calculation method}{\label{calc}}

Our applied quasirelativistic approximation significantly differs from widely
used relativistic Hartree-Fock (HFR) method described in \cite{19}. 
The main differences arise from the set of quasirelativistic Hartree-Fock 
equations (QRHF) \cite{16,17} applied to determine radial orbitals (RO). 
The details of these differences are summarized in \cite{05,06}, while
the most complete and consistent description of our method to calculate
spectroscopic parameters in  the quasirelativistic approach is given in 
\cite{18}. Furthermore, the methods to include effects of the finite nucleus 
size are described in \cite{20,21}. 

The radial orbitals for electrons representing the investigated configurations
were determined by solving the QRHF equations. Initially, the equations were 
solved for the ground configuration $\mathrm{4p^64d^3}$. Afterwards the 
$4f$-electron RO was calculated using the frozen-core potential for the 
$\mathrm{4p^64d^24f}$ configuration. The determined quasirelativistic radial 
orbitals were applied to describe electrons in all configurations. Such an
identical RO basis enables us to avoid the non-orthogonality problems of RO when
the electron transition parameters are calculated. This basis is adopted to 
describe all the quasirelativistic admixed configurations, i.e., the 
configurations obtained from the adjusted configurations by the one-electron 
and two-electron virtual excitations without changing the electron 
principal quantum number ($n=4$, in our case).

The admixed configurations produced by the virtual excitations of one or two 
electrons from the $n=4$ shells to the states with $n>4$ were also included. 
The transformed radial orbitals (TRO) with a variational parameter were applied
to describe these electron states. Initially, the TRO were designed to include 
correlation effects in non-relativistic calculations \cite{22,23,24} and were 
employed successfully for that purpose (see, e.g., \cite{25,26,27,28}). 
The method to determine the TRO in quasirelativistic approximation was presented 
in \cite{18}. In the current work, likewise in \cite{05,06}, we adopt the TRO 
with two variational parameters, the integer even parameter $k$, and the 
positive parameter $\beta$:

\begin{equation}
P_{\mathrm{TRO}}(nl|r) = 
N \left( r^{l-l_0+k} \exp(-\beta r)P_{\mathrm{QR}}(n_0l_0|r) -
\sum_{n^{\prime} < n} P(n^{\prime}l|r) \int_{0}^{\infty} 
P(n^{\prime}l|r^{\prime})r^{\prime\;(l-l_0+k)} \exp(-\beta r^{\prime})
P_{\mathrm{QR}}(n_0l_0|r^{\prime})dr^{\prime} \right).
\end{equation}

Here the factor $N$ ensures the normalization of the determined TRO, the first
term in parenthesis performs the transformation of RO $P_{\mathrm{QR}}$ and the 
second term provides their orthogonality. The parameters $k$ and $\beta$ are 
chosen to gain the maximum of the averaged energy correlation correction 
calculated in the second order of perturbation theory (PT). 

The level energies were calculated in two approximations, A and B. First of all,
likewise in the previous calculations, TRO were determined for the virtual 
electron excitations having the principal quantum number $5 \leq n \leq 7$ and
all allowed values of the orbital quantum number $l$. This is called the 
approximation A. Afterward the TRO basis was extended up to $n \leq 11$. This 
is approximation B. One can generate a huge amount of the admixed configurations
for such bases of RO. For the selection of these configurations, we employed the 
mean weights of the admixed configurations $W_{\mathrm{PT}}$ determined 
in the second  order of PT \cite{29}:

\begin{equation}
W_{\mathrm{PT}} = \frac{\sum_{TLST^{\prime}}(2L+1)(2S+1) 
\left< K_0TLS || H || KT^{\prime} LS \right >^2}
{G(K_0)\left({\bar E}(K) - {\bar E}(K_0)\right)^2}.
\end{equation}

Here the numerator represents the sum of squared sub-matrix elements of the
operators of kinetic energy and electrostatic interaction between the adjusted
configurations $K_0$ and the admixed configurations $K$. The nominator 
represents the statistical weight $G(K_0)$ of adjusted configuration 
multiplied by the squared mean energy differences.

Different selection criteria $W_{\mathrm {min}}$ values for the admixed 
configurations were adopted in case of different sizes of RO bases. We selected
the admixed configurations with $W_{\mathrm{PT}} \geq W_{\mathrm {min}}$.
Here we must emphasize that the number of selected configurations changes only
very slightly if the basis of TRO is extended while the selection parameter
$W_{\mathrm {min}}$ remains the same. The full number of possible configuration
state functions (CSF) was further reduced by permuting the shells of virtually
excited electrons in order to reduce the order of the Hamiltonian being 
diagonalized, as it was described in \cite{30}. This procedure substantially
decreased the number of the $LS$-terms of admixed configurations without any
changes to their correlation corrections.

Our computation abilities are restricted by the amount of memory (RAM) of
computer employed for calculations. In turn, the required amount is determined
by the size of Hamiltonian which depends on the number of CSF having the same
total $LS$ momenta. So this parameter regulates the  $W_{\mathrm {min}}$ values
in our calculations. The computation parameters for two approximations are
given in in Table~\ref{tableA}

\begin{table}[h!]
\caption{\label{tableA}
Computation parameters for two approximations
}
\begin{center}
\begin{tabular}{lrr}
\hline
\multicolumn{1}{c}{Approximation}&
\multicolumn{1}{c}{A} & 
\multicolumn{1}{c}{B}\\
\hline\noalign{\vskip4pt}
Maximum principal quantum number $n$ of TRO                        &        7&                 11\\
Total number of possible interacting even configurations           &      469&               2383\\
Total number of possible even CSF                                  &  1076290&            9591457\\                              
Total number of possible interacting odd configurations            &    10747&               5758\\
Total number of possible odd CSF                                   &  8337946&           78800728\\
Configuration selection parameter  $W_{\mathrm {min}}$             &$10^{-6}$&$1.7 \cdot 10^{-7}$\\
The number of selected interacting even configurations             &      175&                425\\
Complete number of CSF for selected interacting even configurations&   504041&            1782751\\
Reduced number of CSF for selected interacting even configurations &    65168&             115207\\
Maximum number of CSF with the same total $LS$ momenta             &    11417&              20247\\
The number of selected interacting odd configurations              &      357&                894\\
Complete number of CSF for selected interacting odd configurations &  3473850&           13244831\\
Reduced number of CSF for selected interacting odd configurations  &   690686&            1259886\\
Maximum number of CSF with the same total $LS$ momenta             &    62394&             117259\\
\hline\\
\end{tabular}
\end{center}
\end{table}

The energy operator was determined in the quasirelativistic Breit-Pauli 
approximation adopted for the quasirelativistic RO \cite{18}. The calculated 
multiconfigurational eigenvalues and eigenfunctions were employed to compute 
the energy level spectra and to determine data for the electric dipole (E1), 
electric quadrupole (E2), electric octupole (E3) and the magnetic dipole (M1) 
transitions. We have investigated the transitions among the levels of different 
configurations and among the levels of same configurations. Furthermore, the
electron-impact excitation cross-sections and collision strengths were calculated
in the plane-wave-Born approximation with a new computer code described in
\cite{lfz2014} for all considered levels.

To perform our calculations, we have employed our own original computer codes 
together with the codes \cite{31,32,33} which have been adapted for our 
computing needs. The code from \cite{31} was updated according to the methods 
presented in \cite{34,35}.

\section{Results and discussion}{\label{result}}

All the energy levels of the W$^{35+}$ ground configuration 
$\mathrm{4p^64d^3}$ are presented in the review \cite{15} where 
semi-empirical calculations were performed using the experimental data from
\cite{40}. The same data are available from the NIST database \cite{nist}. 
Unfortunately, most of data are graded with uncertain errors ($x$ and $y$), 
classified with question marks ($?$) or only derived from semi-empirical 
calculations. Actually, only three energy levels values relative to the ground 
level are determined accurately. The energy levels of the $\mathrm{4p^64d^3}$ 
configuration together with their $LS$ notations taken from \cite{15} are 
presented in Table~\ref{tableB}. We also present our calculated energy levels 
(approximations A and B) in this table.

\renewcommand{\baselinestretch}{1.0}
\begin{table}[h!]
\caption{\label{tableB}
Comparison of the calculated QR energy levels (in cm$^{-1}$) of W$^{35+}$ 
with the available experimental data from \cite{15}, 
the relativistic MCDF calculation results from \cite{36, 37} 
and the calculation  from \cite{40}. 
}
\begin{center}
\begin{tabular}{rlllrrrrrr}
\hline
$N$ &
\multicolumn{1}{c}{$LS$ \cite{15}}&
\multicolumn{1}{c}{$jj$ \cite{36}} & 
\multicolumn{1}{c}{$J$} & 
\multicolumn{1}{c}{Exp \cite{15, nist}}& 
\multicolumn{1}{c}{QR$_{\mathrm{A}}$}&
\multicolumn{1}{c}{QR$_{\mathrm{B}}$}&
\multicolumn{1}{c}{MCDF \cite{36}} &
\multicolumn{1}{c}{MCDF \cite{37}} &
\multicolumn{1}{c}{HULLAC \cite{40}} \\
\hline\noalign{\vskip4pt}
1		&$^4$F	 & (4d$_-^3$)	            &1.5	&0	      	    &0	     &0	     &0	     &0      &0     \\
2		&	      & (4d$_-^2$)$_2$(4d$_+$)	&2.5	&121554       &122135	&122438	&120830	&120689	&120156\\
3		&$^2$P	 & (4d$_-^2$)$_2$(4d$_+$)	&1.5	&140750       &141621	&141702	&140650	&140288	&140530\\
4		&$^2$G	 & (4d$_-^2$)$_2$(4d$_+$)	&3.5	&156410$+x$  	&153663	&154216	&153100	&152491	&      \\
5		&$^4$P	 & (4d$_-^2$)$_2$(4d$_+$)	&0.5	&160690    	  &157509	&157527	&156620	&156134	&157161\\
6		&$^2$H	 & (4d$_-^2$)$_2$(4d$_+$)	&4.5	&[159500]$+x$ &157228	&157912	&157160	&156200	&      \\
7		&	      & (4d$_-^2$)$_0$(4d$_+$)	&2.5	&[225900]$+y$	&221686	&221938	&221740	&220989	&      \\
8		&$^4$F	 & (4d$_-$)(4d$_+^2$)$_4$ &3.5	&273250$?$   	&274047	&274650	&272150	&271578	&271740\\
9		&	      & (4d$_-$)(4d$_+^2$)$_4$ &4.5	&288570$+x$  	&288100	&288865	&286600	&285712	&      \\
10	&$^4$P	 & (4d$_-$)(4d$_+^2$)$_2$ &1.5	&299520$?$   	&300713	&301018	&298700	&298310	&299234\\
11	&$^2$P	 & (4d$_-$)(4d$_+^2$)$_2$ &0.5	&312200$?$   	&309897	&310060	&308290	&307549	&308555\\
12	&$^2$F	 & (4d$_-$)(4d$_+^2$)$_4$ &2.5	&318120$?$   	&317045	&317427	&315820	&317733	&317389\\
13	&$^2$H	 & (4d$_-$)(4d$_+^2$)$_4$ &5.5	&322010$+x?$ 	&320240	&321357	&319370	&317733	&      \\
14	&$^2$D2	&	(4d$_-$)(4d$_+^2$)$_2$ &2.5	&[344000]     &339160	&339668	&338100	&337135	&      \\
15	&$^2$F	 & (4d$_-$)(4d$_+^2$)$_2$ &3.5	&[350000]     &345289	&345943	&344850	&343433	&      \\
16	&$^2$D1	&	(4d$_-$)(4d$_+^2$)$_0$ &1.5	&402410$+y$  	&399837	&400068	&398540	&397775	&      \\
17	&$^2$G	 & (4d$_+^3$)	            &4.5	&[438000]     &437531	&438446	&435130	&434120	&      \\
18	&$^2$D2	&	(4d$_+^3$)	            &1.5	&[485000]     &479753	&480344	&478100	&476940	&      \\
19	&$^2$D1	&	(4d$_+^3$)	            &2.5	&[516000]     &511846	&512314	&509820	&509005	&      \\
\hline\\
\end{tabular}
\end{center}
\end{table}

Two fully-relativistic calculations of the W$^{35+}$ ion ground configuration 
energy levels are known so far. The first of them is performed in the 
fully-relativistic multiconfiguration Dirac-Fock (MCDF) approach and  has 
included the correlations within the $n = 4$ complex, some 
$n = 4 \rightarrow n^{\prime} = 5$ single excitations and the 
quantum-electrodynamics effects \cite{36}. A very similar MCDF approximation 
involving the CI effects and including the Lamb shift within GRASP2K code is 
exploited in another study \cite{37}. The results of these relativistic 
calculations are presented in Table~\ref{tableB}. The classification of the 
energy levels in the $jj-$coupling is taken from \cite{36}. The experimental 
and theoretical transition wavelengths determined using the relativistic 
HULLAC package are given in \cite{40}. Using those data, we are able to 
reproduce some theoretical energy level values. These parameters are in the last
column of Table~\ref{tableB}.

One can see from Table~\ref{tableB} that substantial extension of CI basis going
from the approximation A to approximation B only slightly alters the energy
level values determined in quasirelativistic approximation. Nevertheless, the
agreement with data from \cite{15} is marginally better for most levels. In 
all cases, the discrepancies from \cite{15} do not exceed $2\%$, and for the
larger part of levels these are smaller than $1\%$. The mean square deviation
\begin{equation}
\sigma = \left(\frac{\sum_i(2J_i+1)(E_i^{[15]}-E_i^{\mathrm{TH}})^2}{\sum_i (2J_i+1)}\right)^{1/2}
\end{equation}
from the values of work \cite{15} goes down from 2696 cm$^{-1}$ to 2355
cm$^{-1}$. On the other hand, both our $\sigma$ values are smaller than those
of calculations by other authors \cite{36, 37}, where $\sigma = 3491$ cm$^{-1}$
and $\sigma = 4296$ cm$^{-1}$, correspondingly. There is one peculiar feature
of theoretical level energies. The deviations from \cite{15} of our QR energy
level values can be negative or positive, whereas both relativistic MCDF 
calculations \cite{36, 37} produce level energies lower than the experimental
data from \cite{15}.

It must be noted that both our QR and other theoretical calculations produce a
different ordering of levels 5 ($J=0.5$) and 6 ($J=4.5$) compared to level
positions given in \cite{15}. When we compare data from \cite{15} with 
the reproduced energy level values from \cite{40}, it is obvious that the 
relative discrepancies are similar to those of other relativistic calculations. 

When compared to the ground configuration levels, the expansion of the CI basis 
affected the calculated levels of the excited configurations 4p$^6$4d$^2$4f and 
4p$^5$4d$^4$ more significantly. Their accuracy has definitely increased. 
Therefore, Table~\ref{table1} contains the calculation results from the 
approximation B. Alongside with level energies, we also provide Lande 
$g$-factors and the level compositions in the $LS$ coupling.

Only one unclassified transition line ($\lambda=53.7$\;\AA) from the excited 
configuration is tabulated in \cite{40}. It has the transition energy 
$\Delta E = 1862000$ cm$^{-1}$. This energy matches very well to the transition 
from the level 142 ($J=1.5$) as one can see from Table \ref{table1}. It has
transition probability to the ground level $A = 2.03 \cdot 10^{12}$ s$^{-1}$.
All other transitions from the level 142 have transition probabilities at least 
by two orders of magnitude smaller. The levels 145 and 147 from 
Table~\ref{table1} have similar transition energies. Their corresponding
transitions probablities are $A = 2.31 \cdot 10^{10}$ s$^{-1}$ and 
$A = 8.68 \cdot 10^{11}$ s$^{-1}$.

Further, using determined CI wavefunction expansions in the quasirelativistic 
approximation, transition parameters for electric multipole (E1, E2 and E3) and
magnetic multipole (M1 and M2) were produced. The high multiple-order 
transitions E3 and M2 were determined following the conclusion made in \cite{41}
where it was demonstrated that those transitions affect calculated radiative 
lifetime values of metastable levels. Our calculated radiative lifetimes
are presented in Table~\ref{table1}.

Since there are no published data for transitions originating from the excited
configurations 4p$^6$4d$^2$4f and 4p$^5$4d$^4$, we can only compare data for
the transitions among the levels of ground configuration 4p$^6$4d$^3$. Analysis
demonstrates that our quasirelativistic calculation results agree very well
with the fully-relativistic calculation results from  \cite{03, 36, 37}. For 
the completeness of spectroscopic data set for the W$^{35+}$ ion, we have also
produced electron-impact excitation parameters, such as collision cross-sections
and collision strengths, for the processes involving the levels of 
configurations 4p$^6$4d$^3$, 4p$^6$4d$^2$4f and 4p$^5$4d$^4$. The database
ADAMANT developed at Vilnius University contains this data set.

The determined theoretical data for the W$^{35+}$ ion can be applied for study
of the excited configurations 4p$^6$4d$^2$4f and 4p$^5$4d$^4$ of this ion or
for the modeling spectra of high-temperature plasma containing tungsten ions.

\ack
The current research is funded by the European Social Fund under the Global 
Grant measure, Project No. VP1-3.1-{\u S}MM-07-K-02-013.

\clearpage

\TableExplanation

\bigskip
\renewcommand{\arraystretch}{1.0}

\section*{Table 1.\label{tab1expl} 
The energy levels $E$ (in cm$^{-1}$), radiative lifetimes $\tau$ 
(in $10^{-9}$s), Lande $g$-factors, and the percentage contributions for 
the energy levels of the $\mathrm{4p^64d^3}$, $\mathrm{4p^64d^24f}$ and  
$\mathrm{4p^54d^4}$ configurations of the W$^{35+}$. 
}

\begin{center}


\end{document}